\documentclass[aps,prl,reprint,twocolumn,showpieces,superscriptaddress]{revtex4-2}
\usepackage{amsmath}
\usepackage{graphicx}
\usepackage{lmodern}
\usepackage{amsmath}
\usepackage{color}
\usepackage{amssymb}
\usepackage{bm}
\usepackage{braket}
\usepackage{mathtools}
\usepackage{afterpage}
\usepackage{soul}
\usepackage[caption=false]{subfig}
\usepackage{ulem}

\usepackage[dvipsnames]{xcolor}
\usepackage[colorlinks=true]{hyperref}
\hypersetup{
  colorlinks=true,
  linkcolor=blue,
  citecolor=blue,
  urlcolor=blue
} 
\makeatletter
\def\maketitle{
\@author@finish
\title@column\titleblock@produce
\suppressfloats[t]}
\makeatother

\begin{document}
\title{\texorpdfstring{Theory of quasiparticle interference in Kitaev quantum spin liquids}{}}
\author{Ammar Jahin}
\affiliation{Theoretical Divison, T-4, Los Alamos National Laboratory, Los Alamos, New Mexico 87545, USA}
\author{Hao Zhang}
\affiliation{Theoretical Divison, T-4, Los Alamos National Laboratory, Los Alamos, New Mexico 87545, USA}
\affiliation{CLNS, Los Alamos National Laboratory, Los Alamos, New Mexico 87545, USA}
\author{Gábor B. Halász}
\affiliation{Materials Science and Technology Divison, Oak Ridge National Laboratory, Oak Ridge, Tennessee 37831, USA}
\affiliation{Quantum Science Center, Oak Ridge, Tennessee 37831, USA}
\author{Shi-Zeng Lin}
\email{szl@lanl.gov}
\affiliation{Theoretical Divison, T-4, Los Alamos National Laboratory, Los Alamos, New Mexico 87545, USA}
\affiliation{CLNS, Los Alamos National Laboratory, Los Alamos, New Mexico 87545, USA}
\affiliation{Center for Integrated Nanotechnologies (CINT), Los Alamos National Laboratory, Los Alamos, New Mexico 87545, USA}

\date{\today}

\begin{abstract}
We study quasiparticle interference (QPI) in the Kitaev quantum spin liquid (QSL) for electrons tunneling into the QSL. The local tunneling conductance around a spin vacancy or localized vison reveals unique features associated with fractionalized Majorana fermions, chargons, and visons. In certain parameter regimes, the single-spinon density of states and momentum dispersion can both be directly extracted from the tunneling conductance. Our results suggest that QPI is a promising tool for identifying the Kitaev QSL and its fractionalized excitations.
\end{abstract} 
\maketitle

\emph{Introduction.}---Quantum spin liquids (QSLs) are exotic phases of matter that avoid long-range ordering at zero temperature~\cite{Balents_2010,Savary_Balents_2017,Zhou_2017,Knolle_2019,Broholm_2020}. 
Rather, these phases are characterized by topological order associated with fractionalization, emergent gauge fields, and Abelian or non-Abelian anyonic quasiparticles. 
The interest in QSLs was initially driven by proposals linking them to high-temperature superconductivity upon doping~\cite{Anderson_1987}. 
Further understanding of QSLs showed that they have promising applications in topological quantum computation~\cite{Kitaev_2003,Nayak_2008,Stern_2013}.
Despite such great interest and potential applications, a definitive experimental observation of these phases is still lacking after decades of search. 

The Kitaev QSL \cite{Kitaev_2006} has recently been the subject of especially intensive study~\cite{Hermanns_2018,Motome_2020,Trebst_2022}. 
In this QSL, the spins are fractionalized into Majorana fermions and $Z_2$ gauge fields. 
Several materials, including (Na,Li)$_2$IrO$_3$~\cite{Jackeli_2009,Singh_2010,Liu_2011,Singh_2012,Chun_2015} and $\alpha$-RuCl$_3$~\cite{Plumb_2014,Sears_2015,Majumder_2015,Kim_2015,Cao_2016} have been proposed to realize the Kitaev QSL with encouraging support from recent experiments. 
In particular, the half-integer-quantized thermal Hall conductivity expected~\cite{Kitaev_2006} for the Kitaev QSL under a magnetic field has been observed~\cite{Kasahara_2018,Yamashita_2020,Yokoi_2021,Bruin_2022}, while evidence of fractionalization is reported from both Raman scattering and inelastic neutron scattering measurements~\cite{Sandilan_2015,Nasu_2016,Banerjee_2016,Banerjee_2017,Do_2017}. Nevertheless, whether the Kitaev QSL is realized in these materials remains debated, calling for more conclusive signatures.

Quasiparticle interference (QPI) around scattering defects is generally a powerful probe of quantum materials~\cite{Avraham_2018,Bena_2008,Brihuega_2008,Mallet_2012,Dombrowski_2017} and emerges as an important tool for studying QSLs~\cite{Wenyu_2022,Ruan_2021,Chen_2022,Kolezhuk_2006,Khaliullin_1997,W_jcik_2023}. 
Indeed, recent scanning tunneling microscopy (STM) measurements on monolayer $\alpha$-RuCl$_3$ have reported distinctive oscillations in the local tunneling conductance around defects~\cite{kohsaka_2024,zheng_2024}. Hence, there is an urgent need to develop a theory for QPI in the Kitaev QSL, which is crucial for both understanding the current experiments and predicting further signatures that can be used to identify Kitaev QSLs in future work.

\begin{figure*}
    \centering
    \captionsetup[subfigure]{
    position=top,
    captionskip=-1pt,
    singlelinecheck=false,
    margin={-2.5cm,1cm}
    }
    \subfloat[]{
    \includegraphics[trim = 77 15 77 5, clip, scale=0.62]{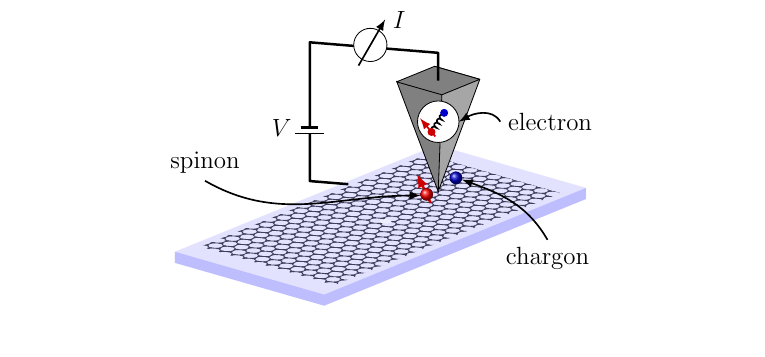}}
    \hfill
    \subfloat[]{\includegraphics[trim = 15 15 12 0, clip,scale=0.62]{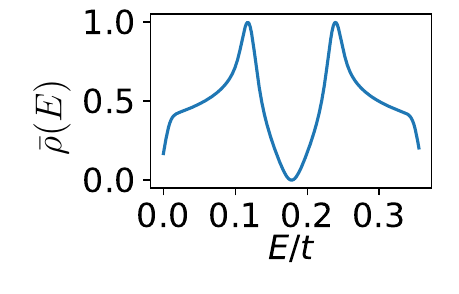}}
    \hfill
    \subfloat[]{\includegraphics[trim = 15 15 12 0, clip,scale=0.62]{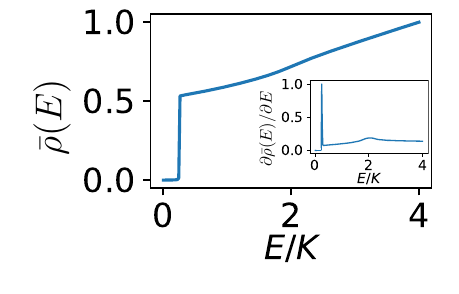}}
    \hfill
    \subfloat[]{\includegraphics[trim = 15 12 8 5, clip,scale=0.62]{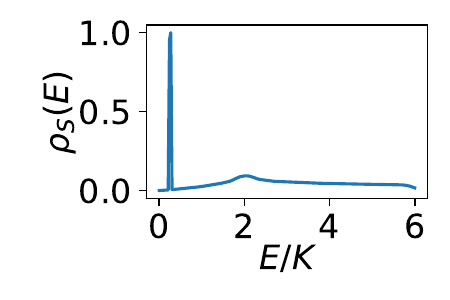}}

    \vspace{-20pt}

    \subfloat[]{\includegraphics[scale=0.8, trim = 0 8 200 5, clip]{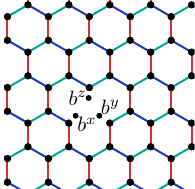}} \hspace{6pt}
    \subfloat[]{\includegraphics[scale=0.42, trim = 140 95 140 68, clip]{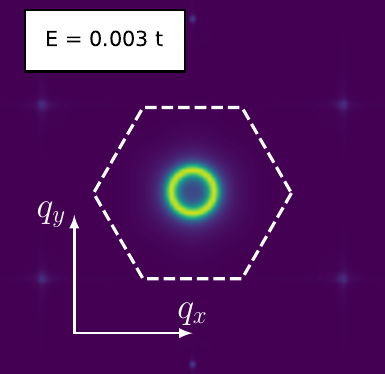}} \hspace{6pt}
    \subfloat[]{\includegraphics[scale=0.42, trim = 140 90 140 75, clip]{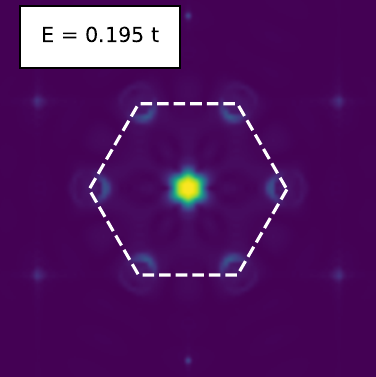}}
    \hfill
    \subfloat[]{\includegraphics[scale=0.8, trim = 200 8 0 5, clip]{defect_defs.pdf}} \hspace{6pt}
    \subfloat[]{\includegraphics[scale=0.42, trim = 140 90 140 75, clip]{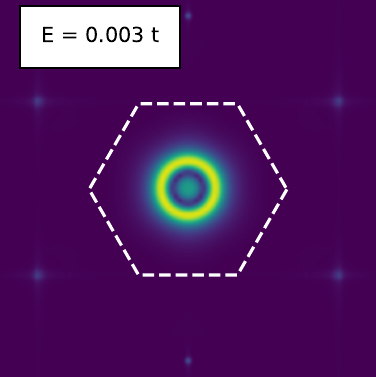}} \hspace{6pt}
    \subfloat[]{\includegraphics[scale=0.42, trim = 140 90 140 75, clip]{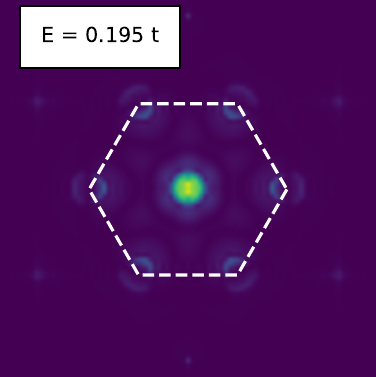}}
    % \subfloat[]{\includegraphics[scale=0.8, trim = 100 0 100 0, clip]{defect_defs.pdf}} \hfill
    % \subfloat[]{\includegraphics[scale=0.437, trim = 140 90 140 75, clip]{chargon_fft_003_b.pdf}} \hfill
    % \subfloat[]{\includegraphics[scale=0.437, trim = 140 90 140 75, clip]{chargon_fft_195_b.pdf}}

    \vspace{-1pt}

    \hfill\includegraphics[scale = 0.6,trim = 0 0 11 103, clip]{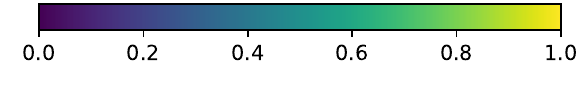} \hspace{93pt}\includegraphics[scale = 0.6,trim = 0 0 11 103, clip]{fig1_colorbar.pdf} \hspace{-10pt}
    \vspace{-16pt}
    
    \caption{(a) Schematic view of an STM experiment performed on a Kitaev QSL. An electron leaving the STM tip is fractionalized into a bosonic chargon carrying the charge and a fermionic spinon carrying the spin. (b-d) STM response of a clean system for fast hopping ($t=10^3K$). (b) Electron DOS at high energies, i.e., well above the Mott gap, follows the chargon DOS showing a graphene-like DOS. (c) Electron DOS (main plot) and its derivative (inset) at low energies, i.e., just above the Mott gap. In this regime, the derivative of the electron DOS follows the spinon DOS. (d) Spinon DOS with a graphene-like component from matter fermions and a sharp peak due to vison pairs. Note that the sharp peak in the spinon DOS corresponds to a step feature in the electron DOS. (e-j) QPI in the STM response around (e-g) a spin vacancy and (h-j) a localized vison for fast hopping ($t=10^3K$). (e,h) Schematic picture of the given defect. (f,g,i,j) Momentum-space electron LDOS around the given defect inside the chargon band where the chargons have (f,i) quadratic and (g,j) Dirac-like dispersion. The white dashed line indicates the first Brillouin zone. In (b) and (c), the electron LDOS is normalized $\bar \rho \equiv (\rho - \rho_{\text{min}}) / (\rho_{\text{max}} - \rho_{\text{min}})$.}
    \label{fig:diff_defects_defs}
\end{figure*}

To realize QPI in the Kitaev QSL, an electron from the STM tip must tunnel directly into the QSL [see Fig.~\ref{fig:diff_defects_defs}(a)], which is only possible when the bias voltage exceeds the Mott gap. This scenario is fundamentally different from previous theoretical setups where electrons were assumed to tunnel through the QSL~\cite{Fledmeier_2020,Elio_2020,Bauer_2023,Takahashi_2023,Kao_2024,Kao_2024_a,Bauer_2024}. In general, the injection of electrons or holes into the Kitaev QSL can give rise to several interesting phenomena such as superconductivity, kinetic ferromagnetism, and fractionalization of electrons, depending on the energy and density of the injected electrons or holes, as well as the details of the microscopic model~\cite{You_2012,Kadow_2024,jin_2023}. In this work, we consider the scenario of spin-charge fractionalization where each electron tunneling into the QSL fractionalizes into a chargon and a spinon [see Fig.~\ref{fig:diff_defects_defs}(a)].  

% \textcolor{red}{[Gabor: Is the Hamiltonian below still a t-K model? I think we need the $\tilde{U}$ term to account for different numbers of charge carriers (0 and 1), i.e., the Mott gap.]}
% \AJ{In this case what is the role of the projector operators? It seems that the projection would be automatically taken care of by the $U$ term, right? Given some fixed filling, the empty sites would cost $U$ energy to create. }

% \textcolor{red}{[Gabor: I think the idea is that the projections get rid of the empty sites (essentially, empty sites have an infinite energy cost) while the $\tilde{U}$ term gives an additional energy for double-occupied sites relative to single-occupied sites (which then roughly corresponds to the Mott gap for electron doping).]}

\emph{Model.}---We use the $t$-$K$ model on a honeycomb lattice near half-filling to describe the Kitaev QSL phase with a doped electron,
\begin{align}
    H =& -K \sum_{\langle \bm r,\bm r'\rangle_{\alpha}} \hat\sigma_{\bm r}^\alpha \hat\sigma_{\bm r'}^\alpha - t \sum_{\langle \bm r,\bm r'\rangle} \sum_{\sigma} \left(\mathcal{P} d^\dagger_{\bm r,\sigma} d^{\phantom{\dagger}}_{\bm r',\sigma} \mathcal{P} + \text{H.c.} \right) \nonumber \\
    &+ \tilde{U} \sum_{\bm r} d^\dagger_{\bm r,\uparrow} d^{\phantom{\dagger}}_{\bm r,\uparrow} d^\dagger_{\bm r,\downarrow} d^{\phantom{\dagger}}_{\bm r,\downarrow},\label{eq:hop_term_elec}
\end{align}
where $d^\dagger_{\bm r,\sigma}$ creates an electron with spin $\sigma$ at site $\bm r$, $\mathcal{P}$ projects onto the single- and double-occupied states at each site, and $\hat{\sigma}_{\bm r}^\alpha = \sum_{i,j} d^\dagger_{\bm r,i} \sigma^\alpha_{ij} d^{\phantom{\dagger}}_{\bm r,j}$ is the spin operator. We describe the fractionalization of electrons with an SU$(2)$ parton mean-field theory where each electron operator is decomposed as $d^\dagger_{\bm r, \sigma} = \frac{1}{\sqrt 2}(a^\dagger_{\bm r, 1} f^\dagger_{\bm r, \sigma} -\eta a^\dagger_{\bm r, 2} f_{\bm r, \bar{\sigma}})$ with $\eta = \pm1$ and $\bar{\sigma}=\,\,\downarrow, \uparrow$ for $\sigma=\,\,\uparrow, \downarrow$, respectively~\cite{You_2012}. The bosonic chargons $a^\dagger_{\bm r, \mu}$ carry the charge of the electron, while the fermionic spinons $f^\dagger_{\bm r, \sigma}$ carry its spin. We assume for concreteness that the Kitaev interaction is ferromagnetic ($K>0$) and that the electron hopping $\propto t$ is spin-independent. We emphasize, however, that our key results do not depend on these assumptions and should apply whenever electron fractionalization takes place.

To find a quadratic problem for the partons $a_{\bm r, \mu}$ and $f_{\bm r, \sigma}$, we use a mean-field decoupling of the hopping term in Eq.~(\ref{eq:hop_term_elec}). The resulting mean-field Hamiltonian is found in the Supplemental Material (SM)~\cite{SM} to be 
\begin{align}\label{eq:mean-field_ham}
    H_{\text{MF}} =& K \sum_{\langle \bm r,\bm r' \rangle_\alpha} u_{\bm r,\bm r'}^\alpha  c_{\bm r} c_{\bm r'} \nonumber \\ 
    &+ \frac{t}{8} \sum_{\langle \bm r,\bm r' \rangle} \sum_{\mu} \left[i W^{\phantom{\dagger}}_{\bm r,\bm r'} a^\dagger_{\bm r, \mu} a^{\phantom{\dagger}}_{\bm r', \mu} + \text{H.c} \right]  \nonumber \\
    &+ \tilde{U} \sum_{\bm r, \mu} a^\dagger_{\bm r, \mu} a^{\phantom{\dagger}}_{\bm r, \mu},
\end{align}
where the spinons are expressed in the Majorana representation as $f_{\bm r, \uparrow} = \frac{1}{2} \left(c_{\bm r}  + i b^z_{\bm r}\right)$ and $f_{\bm r, \downarrow} = \frac{1}{2} \left(ib^x_{\bm r} + b^y_{\bm r}\right)$ with the matter fermions $c_{\bm r}$ and the associated $Z_2$ gauge fields $u_{\bm r, \bm r'}^\alpha = i b^{\alpha}_{\bm r} b^{\alpha}_{\bm r'} = \pm 1$ on the $\alpha = x,y,z$ bonds, as well as the mean-field parameters $W_{\bm r, \bm r'}$. 
% \sout{\AJ{Before discussing our results it is important to comment on the validity of our mean-field treatment. Writing the electronic degrees of freedom in terms of Majorana fermions introduces an SU$(2)$ gauge symmetry to the system. This gauge freedom is broken down to only $Z_2$ for the Kitaev model since the $c_{\bm r}$ Majoranas and the $b_{\bm r}^{\alpha}$ have different dispersions~\cite{You_2012}. The question then becomes if going beyond mean-field gauge fluctuation would be strong enough to confine the chargons and spinons, precluding the possibility of observing these fractionalized particles. We note that since the gauge symmetry is $Z_2$, there is indeed a deconfined phase of the theory with gapped visons in which fractionalization is stable to the small gauge fluctuations~\cite{Senthil_2000}. It is this regime where our results hold, and small gauge fluctuations are not expected to change our results qualitatively. In principle, one can study a strong couple regime where the gauge fluctuations would be strong enough to confine the chargons and spinons, however, this is beyond the scope of our work.}}
In a clean system without any defects, $W_{\bm r, \bm r'} \approx 0.475$ for all bonds, and the chargons thus have an effective hopping strength of $tW_{\bm r, \bm r'}/8 \approx 0.059 t$. The chargon dispersion is then the standard graphene dispersion shifted by energy $\tilde{U}$, with a bandwidth $\Lambda_C = 3tW_{\bm r, \bm r'}/4 \approx 0.356 t$ and a Mott gap $U = \tilde{U} - \Lambda_C/2 > 0$. There are also two Dirac points at finite energy $\tilde{U}$, corresponding to the $\bm K$ and $\bm K'$ points of the Brillouin zone (BZ). The matter-fermion dispersion is the standard graphene dispersion with no energy shift, but is only physical at positive energies, and thus extends down to zero energy with a bandwidth $\Lambda_S = 6K$. Finally, flipping a $Z_2$ gauge field $u_{\bm r, \bm r'}^\alpha$ from $+1$ to $-1$ at any given bond creates a pair of localized gauge fluxes (i.e., visons) at the two neighboring hexagons. These vison pairs form a flat band at constant energy $E_V \approx 0.26K$.

An STM experiment measures the differential tunneling conductance, $dI/dV$, which is proportional to the electron local density of states (LDOS),
 \begin{align}
    \rho(\bm r, E ) = \sum_{m,n,\sigma} |\langle \Psi_{mn} |d^\dagger_{\bm r, \sigma} | \Psi_0 \rangle|^2 \delta(E_{mn} - E),
 \end{align}
where $E = eV - U$ is the bias voltage measured from the Mott gap, and $E_{mn} +  U$ is the energy of the excited state $|\Psi_{mn} \rangle$ relative to the ground state $|\Psi_0 \rangle$. Each state $|\Psi_0 \rangle$ and $|\Psi_{mn} \rangle$ is a product state of a chargon state and a spinon state, $|\Psi_0 \rangle = |\Psi^C_0\rangle \otimes |\Psi^S_0\rangle$ and $|\Psi_{mn} \rangle = |\Psi^C_m\rangle \otimes |\Psi^S_n\rangle$, where $|\Psi^C_0\rangle$ is the chargon vacuum and $|\Psi^S_0\rangle$ is the ground state of the first term in Eq.~(\ref{eq:mean-field_ham}). 
% Since the electron LDOS necessarily vanishes below the chargon (i.e., Mott) gap $U$, we measure the bias voltage from the Mott gap, $E = eV - U$, and introduce a shifted LDOS, $\hat{\rho}(\bm r, E) \equiv \rho(\bm r, E+U)$, 
With our definition of $E$, note that $\rho(\bm r, E)$ is finite down to $E=0$ and is zero for $E< 0$. The electron LDOS can then be written as a convolution,
\begin{align}\label{eq:ldos_conv}
    \rho(\bm r, E) = \int_0^E dE' \rho_C(\bm r, E-E') \rho_S(\bm r, E'),
\end{align}
where $\rho_C(\bm r, E)$ and $\rho_S(\bm r, E)$ are the chargon LDOS and spinon LDOS, respectively, 
\begin{align}
    {\rho}_C(\bm r, E) = \sum_m \sum_\mu |\langle \Psi^C_m| a^\dagger_{\bm r, \mu}| \Psi^C_{0} \rangle|^2 \delta(E_m^C - E),  
\end{align}
\begin{align}
    &\rho_S(\bm r, E) =\sum_n \bigg[|\langle \Psi_n^S |c_{\bm r} | \Psi_0^S \rangle |^2 \nonumber \\ 
     & \qquad \qquad \qquad \quad  \ + \sum_\alpha |\langle \Psi_n^S |b^\alpha_{\bm r} | \Psi_0^S \rangle |^2 \bigg] \delta(E_n^S - E), \label{eq:spinon_ldos}
\end{align}
with $c_{\bm r}$ creating matter fermions and $b^\alpha_{\bm r}$ exciting pairs of visons. Note that $E_m^C$ has the same shift as $E_{mn}$ such that $\rho_C(\bm r, E)$ is finite down to $E=0$.
Details about the valuation of $\rho_C$ and $\rho_S$ can be found in the SM~\cite{SM}, where we use the formalism developed in Refs.~\cite{Bertsch_2012,Carlsson_2021}. 

\emph{Results for fast hopping.}---We first discuss the conductance spectrum at fast hopping, $t\gg K$, {which is the most relevant regime for real materials}. In a clean system, the differential conductance is identical for all sites, and we can thus drop the site index $\bm r$ from each density of states (DOS) in Eqs.~(\ref{eq:ldos_conv})-(\ref{eq:spinon_ldos}). According to Eq.~(\ref{eq:ldos_conv}), the electron DOS ${\rho}(E)$ is then a convolution of a graphene-like chargon DOS ${\rho}_C(E)$ and a spinon DOS $\rho_S(E)$ with two components: a graphene-like contribution coming from matter fermions, and a sharp peak at energy $E_V \approx 0.26K$ corresponding to vison pairs. The full range of ${\rho}(E)$ is plotted in Fig.~\ref{fig:diff_defects_defs}(b); ${\rho}(E)$ follows ${\rho}_C(E)$ with bandwidth $\Lambda_C \approx 0.356 t$ and a graphene-like structure including a linear node in the center and two Van Hove singularities around it. To understand this result, we first notice that $\Lambda_C \gg \Lambda_S$ for $t \gg K$. Hence, for any $E \gg \Lambda_S$, we can approximate the chargon LDOS with a constant in Eq.~(\ref{eq:ldos_conv}) (since $E'<\Lambda_S\ll E$) to obtain
\begin{align}\label{eq:ldos_conv_1}
    {\rho}(\bm r, E) \approx {\rho}_C(\bm r, E) \int_0^{\Lambda_S} dE'  \rho_S(\bm r, E') = 4{\rho}_C(\bm r, E),
\end{align}
where we use a sum rule for the spinon LDOS in the last step. Therefore, we find ${\rho}(E) \sim {\rho}_C(E)$ at $E \gg \Lambda_S$. For $E < \Lambda_S$, in contrast, ${\rho}(E)$ plotted in Fig.~\ref{fig:diff_defects_defs}(c) exhibits a pronounced step at $E \approx E_V \approx 0.26K$. Furthermore, its derivative, $\partial{\rho}(E)/\partial E$ [see inset of Fig.~\ref{fig:diff_defects_defs}(c)], appears to follow $\rho_S(E)$ [see Fig.~\ref{fig:diff_defects_defs}(d)] with a sharp peak at $E \approx E_V$ originating from vison pairs and a much broader peak at $E \approx 2K$ corresponding to the matter-fermion Van Hove singularity. To understand this correspondence between $\partial{\rho}(E)/\partial E$ and $\rho_S(E)$, we differentiate Eq.~(\ref{eq:ldos_conv}),
\begin{align}\label{eq:ldos_conv_2}
    \frac{\partial {\rho}(\bm r,E)}{\partial E} =& \, {\rho}_C(\bm r, E = 0) \rho_S(\bm r, E) \\
    &+ \int_0^E dE' \frac{\partial {\rho}_C(\bm r, E-E')}{\partial E} \rho_S(\bm r, E'), \nonumber
\end{align}
and notice that the first term, $O(1/Kt)$, dominates the second term, $O(1/t^2)$, producing the desired result.

\begin{figure}
    \centering
    \captionsetup[subfigure]{
    position=top,
    captionskip=-2pt,
    singlelinecheck=false,
    margin={-2.5cm,1cm}
    }
    \subfloat[]{\includegraphics[scale=0.41,trim = 11 10 11 5, clip]{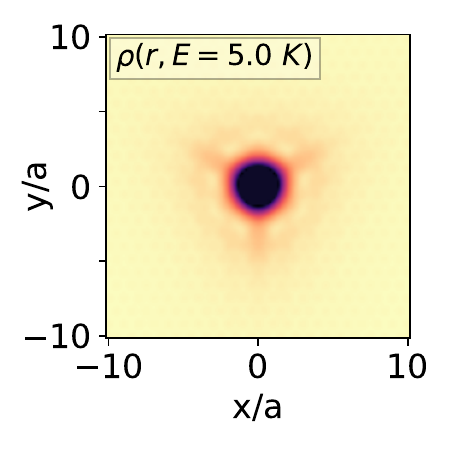}}\hfill
    \subfloat[]{\includegraphics[scale=0.41,trim = 11 10 11 5, clip]{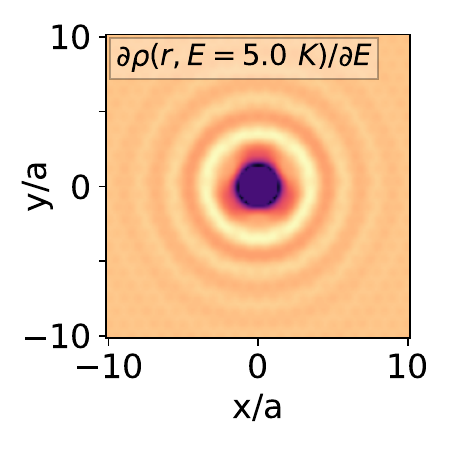}}\hfill
    \subfloat[]{\includegraphics[scale=0.41,trim = 11 10 11 5, clip]{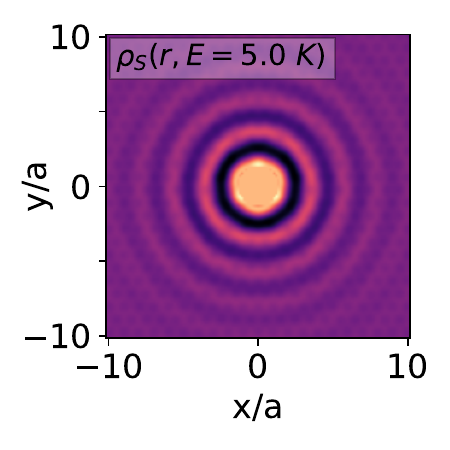}}
    \vspace{-15pt}
    \subfloat[]{\includegraphics[scale=0.45,trim = 10 12 3 5, clip]{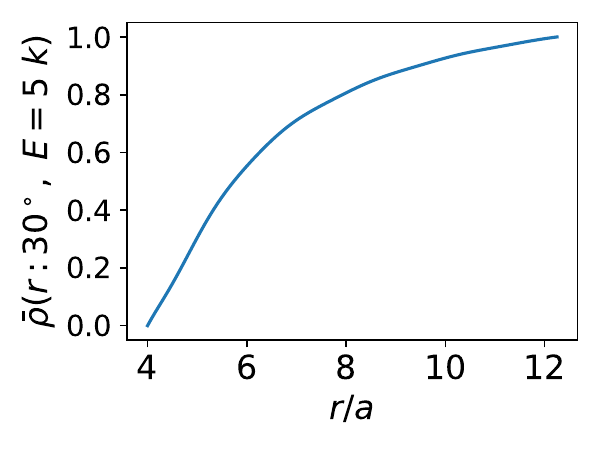}}\hfill
    \subfloat[]{\includegraphics[scale=0.45,trim = 12 12 10 5, clip]{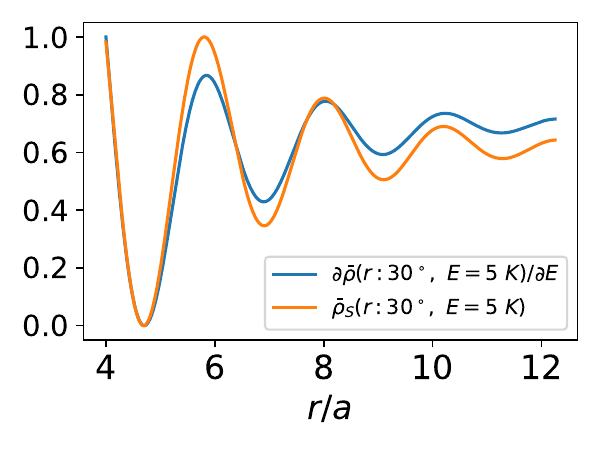}}
    \vspace{-8pt}

    \caption{Low-energy QPI around a spin vacancy for fast hopping ($t=10^4K$). (a,b) Electron LDOS (a) and its derivative (b) around the vacancy. (c) Spinon LDOS around the vacancy. (d) Radial line cut for the electron LDOS. (e) Radial line cut for the derivative of the electron LDOS (blue line) and the spinon LDOS (orange line). While the electron LDOS itself looks featureless, its derivative exhibits clear oscillations following the spinon LDOS.}
    \label{fig:unlock_spinon}
\end{figure}

We now turn to studying the electron LDOS around a defect. We consider two types of naturally occurring defects: a spin vacancy [see Fig.~\ref{fig:diff_defects_defs}(e)], and a localized vison excitation of the QSL with no accompanying structural defect [see Fig.~\ref{fig:diff_defects_defs}(h)]. Like in the clean system, at high bias voltages, $E \gg \Lambda_S = 6K$, the electron LDOS resembles the chargon LDOS. The momentum-space electron LDOS, ${\rho}(\bm q, E) \sim \int d^2 \bm r {\rho}(\bm r, E) e^{-i \bm q \cdot \bm r}$, is shown around a spin vacancy [see Figs.~\ref{fig:diff_defects_defs}(f,g)] and a localized vison [see Figs.~\ref{fig:diff_defects_defs}(i,j)] for two different energies. At the bottom of the chargon band ($E = 0.003t$), the chargon dispersion is quadratic, and the QPI thus exhibits a ring-like feature whose radius is determined by the chargon momentum corresponding to energy $E$ [see Figs.~\ref{fig:diff_defects_defs}(f,i)]. In the middle of the chargon band ($E = 0.195t$), however, the chargons have a Dirac-like dispersion, and the QPI features a filled circle, as opposed to a ring, in the middle of the BZ resulting from the suppressed intra-nodal scattering due to the pseudospin texture [see Figs.~\ref{fig:diff_defects_defs}(g,j)]. There are also arch-looking patterns at the corners of the BZ resulting from the inter-nodal scattering between the $\bm K$ and $\bm K'$ Dirac points. We note that analogous features have been discussed for graphene~\cite{Bena_2008,Bena_2009}. Interestingly, the QPI is similar for a vacancy and a vison; thus, in the absence of any structural defect, the QPI can serve as evidence of a localized vison, which in turn binds a Majorana zero mode under a magnetic field. It is also worth noting that the effect of a vison is felt by the chargons through renormalized hopping amplitudes of the chargons around the vison. The vison is spatially localized in a hexagon and, therefore, produces a QPI similar to a vacancy.

\begin{figure}
    \centering
    \captionsetup[subfigure]{
    position=top,
    captionskip=-2pt,
    singlelinecheck=false,
    margin={-2.5cm,1cm}
    }
    \subfloat[]{\includegraphics[scale=0.46, trim = 7 15 8 8, clip]{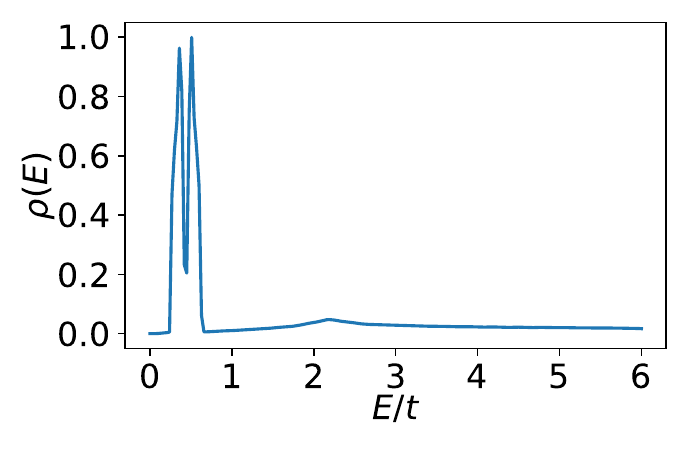}}\hfill
    \subfloat[]{\includegraphics[scale=0.5,  trim = 7 15 10 8, clip]{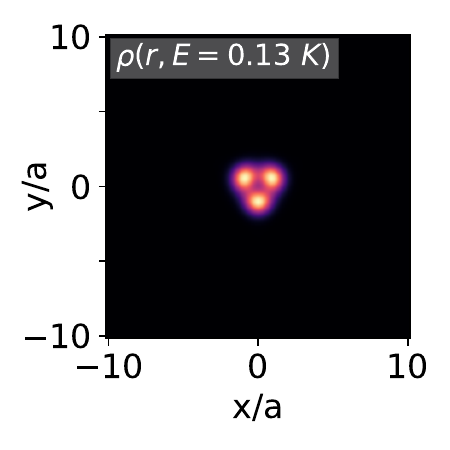}}
    \vspace{-15pt}

    \subfloat[]{\includegraphics[scale=0.5,trim = 7 15 10 8, clip]{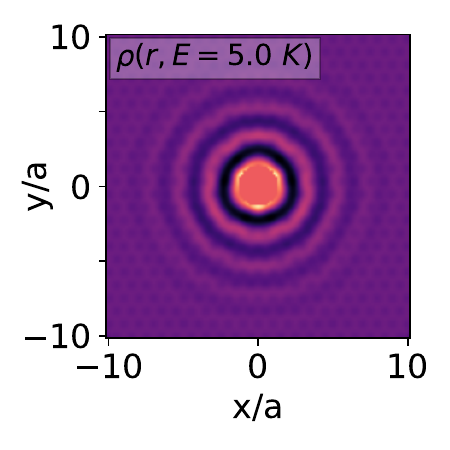}}
    \hspace{15pt}
    \subfloat[]{\includegraphics[scale=0.5,trim = 7 15 10 8, clip]{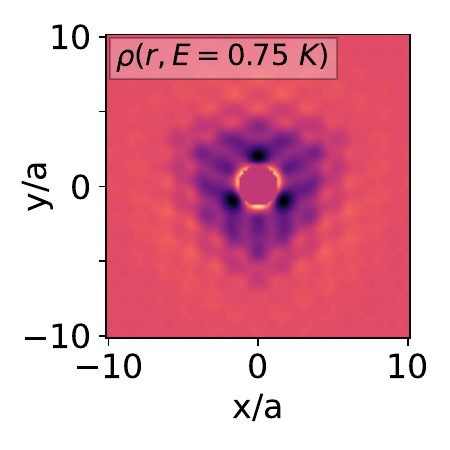}}

    \vspace{-15pt}

    \subfloat[]{\includegraphics[scale=0.5,trim = 7 15 10 8, clip]{spinon_ldos_quadratic.pdf}}
    \hspace{15pt}
    \subfloat[]{\includegraphics[scale=0.5,trim = 7 15 10 8, clip]{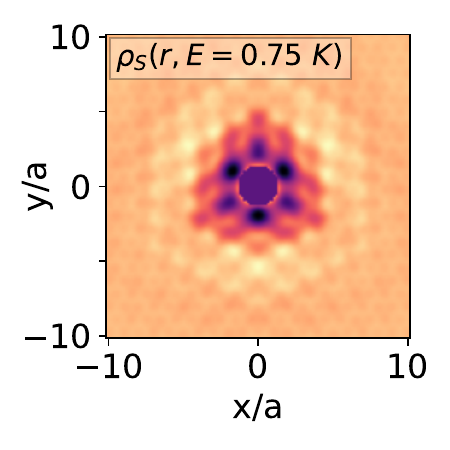}}

    \vspace{-7pt}
    \caption{STM response for slow hopping ($t=K$). (a) Electron DOS for a clean system. (b-f) QPI around a spin vacancy. (b-d) Electron LDOS at $E=0.13K$, $5K$, and $0.75K$, respectively. (e,f) Spinon LDOS at $E = 5K$ and $0.75K$, respectively.}
    \label{fig:teqk}
\end{figure}

Next, we consider the more interesting regime of low bias voltages, $E < \Lambda_S = 6K$, where the spinons play an important role in the electron LDOS. In this regime, the electron LDOS is plotted around a vacancy in Fig.~\ref{fig:unlock_spinon}(a) with a radial line cut shown in Fig.~\ref{fig:unlock_spinon}(d). While the electron LDOS itself appears featureless, its derivative with respect to energy [see Eq.~(\ref{eq:ldos_conv_2})] exhibits distinctive oscillations, as plotted around a vacancy in Fig.~\ref{fig:unlock_spinon}(b) with a radial line cut shown in Fig.~\ref{fig:unlock_spinon}(e). Moreover, as observed by comparing Figs.~\ref{fig:unlock_spinon}(b,e) and Figs.~\ref{fig:unlock_spinon}(c,e), the derivative of the electron LDOS, $\partial {\rho}(\bm r,E)/\partial E$, largely follows the spinon LDOS $\rho_S(\bm r,E)$. This result is readily explained by Eq.~(\ref{eq:ldos_conv_2}) and leads to the remarkable observation that one can directly extract the spinon LDOS from the differential tunneling conductance by taking its derivative with respect to the bias voltage. We further remark that, at very low bias voltages, $E < E_V$, the electron LDOS is mainly contributed from vison pairs near the vacancy, which endows the LDOS with a characteristic three-lobe structure {similar to Fig. \ref{fig:teqk} (b) (Actual results are shown in the SM~\cite{SM})}.

% \SZL{SZL: I do not understand the point of the discussion below. We may delete this part or move them to Supp if it is useful.
% We contrast this by plotting the LDOS at the same energies but without the contribution from the vison bands in Figs.~\ref{fig:effect_vison_bands} (c), and (d). 
% It is clear that including the effects of the vison bands causes most of the weight to be localized near the impurity. 
% This can be understood since the energy cost of exciting visons near the defects is less than in the bulk.
% Take for example the three sites close to a site deletion defect, each has an unpaired $b^\alpha$ Majorana fermion, as shown in Fig.~\ref{fig:diff_defects_defs} (b), thus acting on the ground state with these Majorana operators creates zero energy excitations. 
% As we go farther and farther from the impurity, the cost of exciting a vison increases until it reaches its bulk value of $\approx 0.27K$. 
% These zero-energy vison excitations also exist for a vison defect, though, we expect six zero-energy excitations, instead of three, which correspond to annihilate the original vison defect and create another in one of its six surrounding plaquettes.
% As we create the two vison excitations further and further away from the vertex defect, we also expect this to cost more and more energy, as visons can lower their energy by being close to each other. 
% }

\begin{figure}
    \centering
    \captionsetup[subfigure]{
    position=top,
    captionskip=-2pt,
    singlelinecheck=false,
    margin={-2.5cm,1cm}
    }
    \subfloat[]{\includegraphics[scale = 0.43, trim = 10 15 10 15, clip]{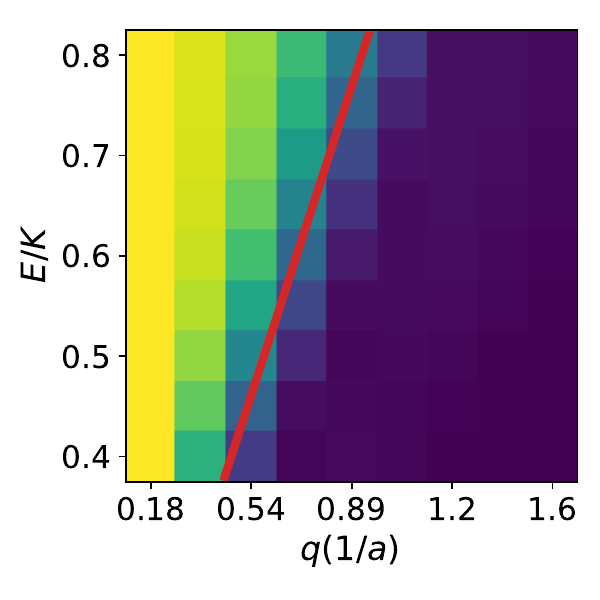}}\hfill
    \subfloat[]{\includegraphics[scale = 0.44,, trim = 10 15 10 15, clip]{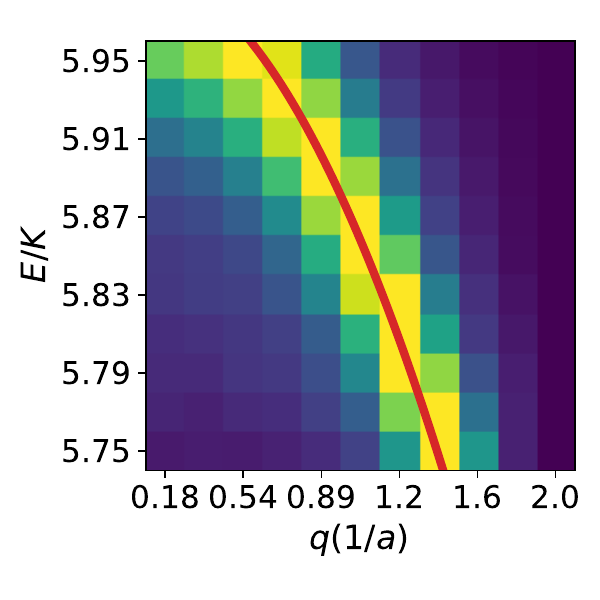}}
    \vspace{-8pt}
    \caption{Extraction of the matter-fermion (i.e., spinon) dispersion for fast hopping ($t=10^4K$). The energy derivative of the electron LDOS, $\partial {\rho}(\bm q, E)/\partial E$, is plotted against both momentum $q=|\bm q|$ and energy $E$ inside the spinon band where the spinons have (a) Dirac-like and (b) quadratic dispersion. (a) The Dirac point with velocity $v$ is reflected in a sharp step at $q = 2E/v$. (b) The spinon dispersion $\mathcal{E}(\bm k)$ can be extracted by tracing the sharp peak at $q = 2\mathcal{E}^{-1} (E)$. In each panel, the red line marks the expected position of the appropriate sharp feature (step or peak) from the Kitaev model.}
    \label{fig:spinon_dispersion}
\end{figure}

We finally extract the matter-fermion dispersion $\mathcal{E}(\bm k)$ by tracing the spinon LDOS, $\rho_S(\bm r, E)\sim\partial \rho(\bm r, E)/\partial E$, in momentum space as the bias voltage $E$ is tuned between $E_V \approx 0.26K$ and $\Lambda_S = 6K$. In Fig.~\ref{fig:spinon_dispersion}(a), $\partial {\rho}(\bm q, E)/\partial E$ is plotted against $q=|\bm q|$ for voltages $E<K$, i.e., close to the Dirac point at the bottom of the matter-fermion band. In this regime, $\partial {\rho}(\bm q, E)/\partial E$ exhibits a filled circle at each voltage $E$ [cf.~Figs.~\ref{fig:diff_defects_defs}(g,j)] with a sharp step at its radius, $q_{\mathrm{QPI}}=2 E / v$, where $v$ is the matter-fermion velocity at the Dirac point. Hence, this step feature confirms the existence of a matter-fermion Dirac point in the Kitaev QSL. In Fig.~\ref{fig:spinon_dispersion}(b), $\partial {\rho}(\bm q, E)/\partial E$ is plotted at the top of the matter-fermion band where the dispersion is quadratic. In this regime, $\partial {\rho}(\bm q, E)/\partial E$ possesses a sharp ring-like feature at each voltage $E$ [cf.~Figs.~\ref{fig:diff_defects_defs}(f,i)], with its peak radius $q_{\mathrm{QPI}}$ being twice the matter-fermion momentum corresponding to energy $E$. Indeed, the ring-like feature in Fig.~\ref{fig:spinon_dispersion}(b) follows $E = \mathcal{E}(\bm q/2)$ and can thus be exploited to directly extract the matter-fermion dispersion $\mathcal{E}(\bm k)$ of the Kitaev QSL.  

\emph{Results for slow hopping.}---In the regime of slow hopping, $t \lesssim K$, the chargon bandwidth is much smaller than the spinon bandwidth, $\Lambda_C\ll\Lambda_S$, because $\Lambda_C\approx0.356t$ and $\Lambda_S=6K$. Thus, for any bias voltage $E\gg\Lambda_C$, one may use an approximation analogous to Eq.~(\ref{eq:ldos_conv_1}), but with interchanged roles of the spinons and the chargons, to argue that the electron LDOS ${\rho}(\bm r, E)$ largely follows the spinon LDOS $\rho_S(\bm r, E)$. The only caveat is that $\rho_S(\bm r, E)$ is not smooth at the vison energy, $E_V\approx0.26K$, which restricts this approximation to $E\gg\max(\Lambda_C,E_V)$. Indeed, while the electron DOS of a clean system shown in Fig.~\ref{fig:teqk}(a) resembles the spinon DOS [see Fig.~\ref{fig:diff_defects_defs}(d)] for $E\gtrsim K$, with a broad peak at $E\approx2K$ due to the matter-fermion Van Hove singularity, the sharp peak at $E\approx E_V$ coming from vison pairs is doubled as it is convolved with the bimodal chargon DOS [see Fig.~\ref{fig:diff_defects_defs}(b)]. {In the presence of vacancy, for $E$ near $E_V$, the contribution to the electron LDOS near a vacancy is dominated by the flat vison bands, which yields well localized three-lobe structure as shown in Fig. \ref{fig:teqk} (b). When $E \gg \max(\Lambda_C,E_V)$, electron LDOS [Fig.~\ref{fig:teqk}(c)] shows clear oscillations following those of the spinon LDOS [Fig.~\ref{fig:teqk}(e)]. This result suggests an even more direct correspondence between the QPI and the matter-fermion dispersion than in the previous regime of fast hopping. For $E \sim \max(\Lambda_C,E_V)$, there is no clear separation of energy scale between chargon and spinon, because the bias voltage is not enough to access the full spinon band. Nevertheless, clear spatial oscillation of electron LDOS occurs [Fig.~\ref{fig:teqk} (d)] although a direct correspondence to the directly calculated spinon LDOS [Fig.~\ref{fig:teqk} (f)] is less obvious. }
%Nevertheless, the electron LDOS near a vacancy for $E\gtrsim K$, as depicted in Fig.~\ref{fig:teqk}(c,d), shows clear oscillations that appear to follow those of the spinon LDOS [see Fig.~\ref{fig:teqk}(e,f)]. This result suggests an even more direct correspondence between the QPI and the matter-fermion dispersion than in the previous regime of fast hopping. 
%When $E-W_c>E_V$, the electron LDOS is given by the convolution of spinon LDOS and chargon LDOS, which are no longer separated, Figs.~\ref{fig:teqk} (c) and (d). In the case when $E \approx E_V$, the electron LDOS is mainly contributed from the localized vison bands [See Fig.~\ref{fig:teqk} (b)].

\emph{Conclusions}--- We study QPI in a Kitaev QSL next to an impurity, as directly measureable in an STM experiment.
Employing a parton mean-field theory, we obtain the electron LDOS as a convolution of the chargon and spinon LDOS, and highlight several distinctive features that can help identify the Kitaev QSL.
First, we point out that the electron DOS of a clean system exhibits signatures of both vison pairs at $E\approx E_V\approx0.26K$ and a matter-fermion (i.e., spinon) Van Hove singularity at $E\approx2K$.
%The effects of the vison bands also show in the LDOS at $E < E_v$, where the LDOS weight is localized to only a few lattice constants near the vacancy.
Second, in the presence of an impurity, we describe a simple protocol to extract the spinon LDOS and, hence, the spinon dispersion from the measured electron LDOS, thus providing direct evidence of a spinon Dirac point with linear dispersion at low energy.
This protocol is remarkable because it allows us to access key spinon properties that are otherwise hard to measure due to the fractionalized nature of the spinons.
We also emphasize that the QPI exhibits similar behavior near structural defects like spin vacancies and localized vison excitations that do not alter the structure but act like impurities for spinons.
Our results indicate that QPI is a powerful technique for identifying both the Kitaev QSL and its fractionalized excitations.
Moreover, our approach readily generalizes to other types of QSLs, such as U(1) QSLs with spinon Dirac points or Fermi surfaces.

{
We finally discuss the role of additional magnetic interactions commonly present in Kitaev magnets, such as Heisenberg and Gamma interactions. In the presence of such non-Kitaev interactions, the Kitaev QSL remains stable within a finite region of the parameter space~\cite{PhysRevLett.112.077204}, and a straightforward generalization of our parton approach (see the SM~\cite{SM}) can still be used to calculate the QPI inside the Kitaev QSL phase. Through a renormalization of the parton band structure, non-Kitaev interactions lead to quantitative (but not qualitative) changes in the QPI. Another notable effect of non-Kitaev interactions is the enhancement of $Z_2$ gauge fluctuations. Deep inside the Kitaev QSL phase, the visons are gapped, resulting in weak gauge fluctuations. However, as the system approaches a phase transition out of the Kitaev QSL, gauge fluctuations become more pronounced. Analogous to electrons interacting with fluctuating gauge fields~\cite{Sachdev_2011}, these enhanced fluctuations are expected to dampen the QPI signal and may eventually smear it out, particularly at higher energies where gauge fluctuations are stronger. Consequently, QPI serves as a powerful tool for probing not only the Kitaev QSL itself but also the phase transitions that separate it from nearby phases.}

\begin{acknowledgements}
{\it Acknowledgments.---} We thank Yuhki Kohsaka, Yuji Matsuda, Cristian Batista, and Natalia Perkins for helpful discussions, and particularly Yuji Matsuda for sharing their unpublished manuscript. The work at LANL was carried out under the auspices of the U.S. DOE NNSA under contract No. 89233218CNA000001 through the LDRD Program, and was supported by the Center for Nonlinear Studies at LANL, and was performed, in part, at the Center for Integrated Nanotechnologies, an Office of Science User Facility operated for the U.S. DOE Office of Science, under user proposals $\#2018BU0010$ and $\#2018BU0083$. G.B.H.~was supported by the U.S.~Department of Energy, Office of Science, National Quantum Information Science Research Centers, Quantum Science Center.

{This manuscript has been authored by UT-Battelle, LLC, under contract DE-AC05-00OR22725 with the US Department of Energy (DOE). The publisher acknowledges the US government license to provide public access under the DOE Public Access Plan (http://energy.gov/downloads/doe-public-access-plan).}
\end{acknowledgements}

\bibliography{references}

{
\setcounter{equation}{0}
\clearpage

\setcounter{figure}{0}
\setcounter{table}{0}

\onecolumngrid
\appendix

\title{\texorpdfstring{Supplementary materials for ``theory of quasiparticle interference in Kitaev quantum spin liquids"}{}}

\section{Parton decomposition and evaluation of the mean-field parameters}
In this section, we give the details for the parton construction and the mean-field theory Hamiltonian we use in the main text. 
The Hamiltonian governing the electron hopping is, 
\begin{align}
    H_t = -t \sum_{\langle \bm r,\bm r'\rangle}\sum_{\sigma}\left[\mathcal{P} d^\dagger_{\bm r,\sigma} d_{\bm r',\sigma} \mathcal{P} + \text{h.c.} \right].\label{eq:hop_term_elec}
\end{align}
where $\mathcal{P}$ projects out the local empty states. 
We express the local electronic degrees of freedom in terms of two flavors of chargons $a_{\bm r, \mu}$ and two flavors of spinons $f_{\bm r, \sigma}$
\begin{align}\label{eq:parton_construction}
    d^\dagger_{\bm r, \uparrow} = \frac{1}{\sqrt 2}(a^\dagger_{\bm r, 1} f^\dagger_{\bm r, \uparrow} - a^\dagger_{\bm r, 2} f_{\bm r, \downarrow}) && 
    d^\dagger_{\bm r, \downarrow} = \frac{1}{\sqrt 2}(a^\dagger_{\bm r, 1} f^\dagger_{\bm r, \downarrow} + a^\dagger_{\bm r, 2} f_{\bm r, \uparrow}), 
\end{align}
Such construction introduces an SU$(2)$ gauge freedom, which is best seen rewriting the above equations as, 
\begin{align}
    \mathcal D_{\bm r} = \frac{1}{\sqrt{2}} \mathcal F_{\bm r} \mathcal A_{\bm r}
\end{align}
with, 
\begin{align}\label{eq:mat_form_defs}
    \mathcal D_{\bm r} = \begin{bmatrix}
                    d^{\dagger}_{\bm r, \uparrow} & -d_{\bm r, \downarrow} \\ 
                    d^\dagger_{\bm r, \downarrow} & d_{\bm r, \uparrow}
                \end{bmatrix}, && 
    \mathcal F_{\bm r} = \begin{bmatrix}
                    f^\dagger_{\bm r, \uparrow} & -f_{\bm r, \downarrow} \\ 
                    f^\dagger_{\bm r, \downarrow} & f_{\bm r, \uparrow}
                \end{bmatrix}, && 
    \mathcal A_{\bm r} = \begin{bmatrix}
                    a^\dagger_{\bm r, 1} & -a_{\bm r, 2} \\ 
                    a^\dagger_{\bm r, 2} & a_{\bm r, 1}
                \end{bmatrix}.
\end{align}
In this form the spin SU$(2)$ rotations are implemented by $\mathcal D_{\bm r} \rightarrow U_s \mathcal D_{\bm r}$. 
In addition, this expansion of the Hilbert space introduces a gauge freedom in that the physical degrees of freedom $\mathcal D_{\bm r}$ do not change under $\mathcal F_{\bm r} \rightarrow \mathcal{F}_{\bm r} U_{g}$ and, $\mathcal A_{\bm r} \rightarrow U^\dagger_{g}\mathcal{A}_{\bm r} $. 
The generators of this SU$(2)$ gauge symmetry are, 
\begin{align}
    K^{\alpha}_{\bm r} = \frac{1}{4} \text{Tr}\  \mathcal F_{\bm r}\sigma^{\alpha}F^\dagger_{\bm r} - \frac{1}{4} \text{Tr} \  \sigma^z \mathcal A^\dagger_{\bm r}\sigma^{\alpha}A_{\bm r}
\end{align}
The physical Hilbert space is invariant under such gauge transformation, i.e. $K^\alpha_{\bm r} = 0$. 
It can be shown that only three states satisfy this condition, 
\begin{align}
    d^\dagger_{\bm r, \uparrow} |0\rangle = f^\dagger_{\bm r, \uparrow} |\tilde 0 \rangle, && d^\dagger_{\bm r, \downarrow} |0\rangle = f^\dagger_{\bm r, \downarrow} |\tilde 0 \rangle, && d^\dagger_{\bm r, \downarrow} d^\dagger_{\bm r, \uparrow} |0\rangle = \frac{1}{\sqrt{2}} \left( a^\dagger_{\bm r, 2} +a^\dagger_{\bm r, 1}f^\dagger_{\bm r, \downarrow}f^\dagger_{\bm r, \uparrow} \right) |\tilde 0 \rangle. 
\end{align}
where $d_{\bm r, \sigma}|0 \rangle = 0$ and $f_{\bm r, \sigma} |\tilde 0 \rangle = a_{\bm r, \mu} |\tilde 0 \rangle = 0$. 
We choose the decomposition of the electronic degrees of freedom into spinons and chargon so that the physical space coincides with the low energy subspace formed by the single, and doubly occupied local states in the presence of electron doping.

Substituting Eq.~\eqref{eq:parton_construction} in $H_t$ we get, 
\begin{align}
    H_t = -t \sum_{\langle \bm r,\bm r'\rangle} \left[ a^\dagger_{\bm r, 1} a_{\bm r', 1} \bigg( f^\dagger_{\bm r ,\uparrow} f_{\bm r', \uparrow} +   f^\dagger_{\bm r, \downarrow} f_{\bm r', \downarrow}  \bigg) + a^\dagger_{\bm r, 2} a_{\bm r', 2}  \bigg(f_{\bm r, \uparrow} f^\dagger_{\bm r', \uparrow} + f_{\bm r, \downarrow} f^\dagger_{\bm r', \downarrow}\bigg)  \right. \nonumber \\ 
    \left. + a^\dagger_{\bm r, 1} a_{\bm r', 2} \bigg(f^\dagger_{\bm r, \downarrow}f^\dagger_{\bm r', \uparrow} -  f^\dagger_{\bm r,\uparrow }f^\dagger_{\bm r', \downarrow} \bigg) + a^\dagger_{\bm r, 2} a_{\bm r', 1} \bigg(f_{\bm r, \uparrow}f_{\bm r', \downarrow} -  f_{\bm r, \downarrow}f_{\bm r',\uparrow } \bigg) +   \text{h.c.} \right].
\end{align}
We use a mean-field theory to study this Hamiltonian. We consider the fractionalized regime, where the mean-field order parameters for the $f$ bilinears are those of the Kitaev quantum spin liquid.  
The ansatz takes the following form, 
\begin{align}
    H^{\text{MF}}_t = \frac{t}{8} \sum_{\bm r,\bm r' }   \sum_{\mu} i W_{\bm r,\bm r'} a^\dagger_{\bm r, \mu} a_{\bm r', \mu}
\end{align}
where
\begin{align}
    \langle a^\dagger_{\bm r, \mu} a_{\bm r', \mu'} \rangle =   \langle f^\dagger_{\bm r, \uparrow} f^\dagger_{\bm r', \downarrow'} \rangle = 0
\end{align}
and
\begin{align}
    W_{\bm r,\bm r'} = \langle ic_{\bm r} c_{\bm r'} + \sum_\alpha i b^\alpha_{\bm r} b^\alpha_{\bm r'} \rangle,
\end{align}
where we have defined the Majorana operators, 
\begin{align}
    f^\dagger_{\bm r, \uparrow} = \frac{1}{{2}} (c_{\bm r} + i b^z_{\bm r}), && f^\dagger_{\bm r, \downarrow} = \frac{1}{{2}} (ib^x_{\bm r} +  b^y_{\bm r}). 
\end{align}
The expectation values $W_{\bm r, \bm r'}$ are evaluated for the ground state of the Kitaev model with a defect introduced. This requires solving the quadratic Hamiltonian of the Kitaev model in the presence of a defect which can be done numerically. 

The terms $\langle i b^\alpha_{\bm r} b^\alpha_{\bm r'} \rangle$ are easy to compute. They are nonzero only when the product $ b^\alpha_{\bm r} b^\alpha_{\bm r'}$ does not create any visons. 
This happens when $\bm r$ and $\bm r'$ are nearest neighbors and make an $\alpha$-edge, 
\begin{align}
    \langle ib^\alpha_{\bm r}b^\alpha_{\bm r'} \rangle = \begin{cases}
        u^{std}_{\bm r, \bm r'} & \bm r \text{ and } \bm r' \text{ make an } \alpha \text{-edge}  \\ 
        0 & \text{otherwise}
    \end{cases}
\end{align}
where $u^{std}_{\bm r, \bm r'}$ is the standard gauge field defining the vison-free subspace away from a defect and zero for edges connected to a vacancy.  

The evaluation of $\langle ic_{\bm r} c_{\bm r'} \rangle $ requires the diagonalization of the Kitaev Hamiltonian which takes the general form 
\begin{align}
    H_K = \frac{K}{2} \sum_{\bm r, \bm r'} A_{\bm r, \bm r'} \ i c_{\bm r} c_{\bm r'}
\end{align} 
Since $A_{\bm r, \bm r'}$ is an antisymmetric matrix, there exists a matrix $Q \in O(N)$, with $N$ being the number of sites of the system, such that 
\begin{align}\label{eq:off_diag_decomp}
    A = Q \mathcal E Q^T
\end{align}
with 
\begin{align}
    \mathcal E = \bigoplus^{N/2}_{m = 1} \begin{bmatrix}
        0 & \varepsilon_m \\ 
        -\varepsilon_m & 0
    \end{bmatrix}, \qquad \varepsilon_m > 0.
\end{align}
Using this to rewrite the Kitaev Hamiltonian, we have 
\begin{align}\label{eq:diag_kitaev_ham}
    H_K = \sum_{m = 1}^{N/2} \varepsilon_m \ i b_{2m} b_{2m + 1} = \sum_m \varepsilon_m \left[ 2 \psi_m^\dagger \psi_m  - 1 \right]
\end{align}
where we define $b_i = c_{\bm r} Q_{\bm r, i}$, and $\psi_m = (b_{2m} + i b_{2m + 1})/2$.
We obtain the matrix $Q$ numerically which has a computational cost of $O(N^3)$, the same as the exact diagonalization.
The ground state of the Kitaev model is defined such that $\psi_m |\Psi_0^S\rangle = 0$ for all $m$. 
In the new basis, we can easily evaluate, 
\begin{align}
    \langle ib_{2m} b_{2n}  \rangle = \langle ib_{2m + 1} b_{2n + 1}  \rangle =  i \delta_{m,n}, && \langle ib_{2m} b_{2n + 1}  \rangle  = - \delta_{m,n}
\end{align}
Using this we can write
\begin{align}
    \langle ic_i c_j \rangle = Q_{j,j'} Q_{i,i'} \langle ib_{i'} b_{j'} \rangle 
\end{align}
which we evaluate numerically.

\section{Parton construction of generic Hamiltonians}
A Hamiltonian for a family of honeycomb lattice spin liquid candidate materials {can be} written as
\begin{equation}
    \mathcal{H} = \sum_{\langle \bm r, \bm r' \rangle \in \alpha\beta(\gamma)} \big[J\bm{S}_{\bm r}\cdot\bm{S}_{\bm r'}+KS_{\bm r}^{\gamma}S_{\bm r'}^{\gamma}+\Gamma(S_{\bm r}^{\alpha}S_{\bm r'}^{\beta}+S_{\bm r}^{\beta}S_{\bm r'}^{\alpha})\big] -  \sum_{\bm r} \bm B \cdot \bm S_{\bm r},
    \label{eq:generic_ham}
\end{equation}
where $J$ is the Heisenberg exchange, $K$ is the Kitaev exchange, and $\Gamma$ denotes the symmetric off-diagonal exchange, and $\bm B$ represents an external Zeeman field. We introduce the parton (complex) fermions to represent the local moment $\bm{S}_{\bm r}=\frac{1}{4}\text{Tr} \mathcal F_{\bm r}^{\dagger}\bm{\sigma}\mathcal F_{\bm r}$, where the matrix $\mathcal F_{\bm r}$ is as defined in Eq.~\eqref{eq:mat_form_defs}, and we impose the constraint $f_{\bm r \alpha}^{\dagger}f_{\bm r \alpha}=1$, {where, for here and the rest of this section, repeated Greek indices are summed over}. 
The Zeeman term takes the following quadratic form in fermionic representation, 
\begin{align}
    \bm B \cdot \bm S_{\bm r} =  \frac{1}{2}B^a \sigma^a_{\alpha, \beta} f^\dagger_{\bm r, \alpha} f_{\bm r, \beta} 
\end{align}
{The magnetic interactions $J$, $K$, and $\Gamma$ are all quartic in the fermionic language and take the following form
\begin{align}
    S^{a}_{\bm r} S^{a'}_{\bm r'} = \frac{1}{4} \left(f^\dagger_{\bm r \alpha} \sigma^{a}_{\alpha, \beta} f_{\bm r, \beta}  \right) \left(f^\dagger_{\bm r' \alpha'} \sigma^{a'}_{\alpha', \beta'} f_{\bm r', \beta'}  \right), \qquad a, a' \in \{0,x,y,z\}
\end{align}
We can treat such quartic interactions using a mean-field theory. To facilitate a mean-field treatment, we further write this interaction in the following way, 
\begin{align}
     S^{a}_{\bm r} S^{a'}_{\bm r'} = \frac{1}{8} \left[  (f_{\bm r \alpha} f_{\bm r' \alpha'})^\dagger  \sigma^{a}_{\alpha, \beta}   \sigma^{a'}_{\alpha', \beta'} f_{\bm r, \beta}   f_{\bm r', \beta'}  \right] - \frac{1}{8} \left[ (f^\dagger_{\bm r, \beta} f_{\bm r' \alpha'})^\dagger \sigma^{a}_{\alpha, \beta}  \sigma^{a'}_{\alpha', \beta'}  f^\dagger_{\bm r \alpha} f_{\bm r', \beta'}  \right].
\end{align}
}
We introduce the following fermion bilinears,
\begin{equation}
    % \hat{\rho}^0_{\bm r, \bm r'} = f_{\bm r\alpha}^{\dagger}f_{\bm r'\alpha}, \ \
    % \hat{{\Delta}}^0_{\bm r, \bm r'} = f_{\bm r\alpha}(i\tau^y)_{\alpha\beta} f_{\bm r'\beta}, \ \
    \hat{\rho}^{a}_{\bm r, \bm r'} = f_{\bm r \alpha}^{\dagger}(\sigma^a)_{\alpha\beta}f_{\bm r'\beta}, \ \
   {\hat{\Delta}^{a}_{\bm r, \bm r'} = f_{\bm r \alpha}(i\sigma^y\sigma^a)_{\alpha\beta}f_{\bm r'\beta},} \ \ \qquad a \in \{0,x,y,z\}
\end{equation}
where $\hat{\rho}^{s}_{\bm r, \bm r'}$ and $\hat{{\Delta}}^{s}_{\bm r, \bm r'}$ are SU(2) {spin} singlet and $\hat{\rho}^{a}_{\bm r, \bm r'}$ and $\hat{\Delta}^{a}_{\bm r, \bm r'}$ are SU(2) {spin} triplet with $a=x,y,z$.
{We can write bilinears $f_{\bm r\alpha} f_{\bm r', \beta}$ in terms of the $\hat{\rho}$ and $\hat \Delta$ using
\begin{align}
    f_{\bm r, \uparrow} f_{\bm r', \uparrow} = \frac{-1}{2} (\hat \Delta^{x}_{\bm r, \bm r'} + i \hat \Delta^y_{\bm r, \bm r'} ), && f_{\bm r, \downarrow} f_{\bm r', \downarrow} = \frac{1}{2} (\hat \Delta^{x}_{\bm r, \bm r'} - i \hat \Delta^y_{\bm r, \bm r'} ), \nonumber \\  f_{\bm r, \uparrow} f_{\bm r', \downarrow} = \frac{1}{2}(\hat \Delta_{\bm r, \bm r'} + \hat \Delta^z_{\bm r, \bm r'}), && f_{\bm r, \downarrow} f_{\bm r', \uparrow} = \frac{1}{2}(-\hat \Delta_{\bm r, \bm r'} + \hat \Delta^z_{\bm r, \bm r'})
\end{align}
and 
\begin{align}
     f^\dagger_{\bm r, \uparrow} f_{\bm r', \uparrow} = \frac{1}{2}(\hat \rho_{\bm r, \bm r'} + \hat \rho^z_{\bm r, \bm r'} ) && f^\dagger_{\bm r, \downarrow} f_{\bm r', \downarrow} = \frac{1}{2}(\hat \rho_{\bm r, \bm r'} - \hat \rho^z_{\bm r, \bm r'} ) \nonumber \\ 
     f^\dagger_{\bm r, \uparrow} f_{\bm r', \downarrow} = \frac{1}{2}(\hat \rho^x_{\bm r, \bm r'} - i\hat \rho^y_{\bm r, \bm r'} ) && f^\dagger_{\bm r, \downarrow} f_{\bm r', \uparrow} = \frac{1}{2}(\hat \rho^{x}_{\bm r, \bm r'} + i\hat \rho^y_{\bm r, \bm r'} ). 
\end{align}
Using these relationships we can work out the expressions for the Kitaev interactions, 
\begin{align}
    K S^x_{\bm r} S^{x}_{\bm r'} = \frac{-K}{16} \left[ \hat \Delta_{\bm r, \bm r'}^{0\dagger}\hat \Delta^0_{\bm r, \bm r'} + \hat \Delta^{x^\dagger}_{\bm r, \bm r'} \hat \Delta^x_{\bm r, \bm r'} -  \hat \Delta^{y^\dagger}_{\bm r, \bm r'} \hat \Delta^y_{\bm r, \bm r'} -  \hat \Delta^{z^\dagger}_{\bm r, \bm r'} \hat \Delta^z_{\bm r, \bm r'}  + \hat \rho_{\bm r, \bm r'}^{0\dagger}\hat \rho^0_{\bm r, \bm r'} + \hat \rho^{x^\dagger}_{\bm r, \bm r'} \hat \rho^x_{\bm r, \bm r'} -  \hat \rho^{y^\dagger}_{\bm r, \bm r'} \hat \rho^y_{\bm r, \bm r'} -  \hat \rho^{z^\dagger}_{\bm r, \bm r'} \hat \rho^z_{\bm r, \bm r'}    \right]. 
\end{align}
Expressions for $S^y_{\bm r}S^y_{\bm r'}$ and $S^z_{\bm r}S^z_{\bm r'}$ can be worked out by permitting $x,y,z$ in a cyclic manner, $x,y,z \rightarrow y,z,x$.}

{
By adding $S^x_{\bm r}S^x_{\bm r'}+S^y_{\bm r}S^y_{\bm r'}+S^z_{\bm r}S^z_{\bm r'}$ we obtain the expression for the 
} 
Heisenberg exchange term 
\begin{equation}
    J \bm{S}_{\bm r}\cdot\bm{S}_{\bm r} = -\frac{3J}{16} \left[ \left(\hat{\rho}^{0}_{\bm r, \bm r'} \right)^{\dagger}\hat{\rho}^{0}_{\bm r, \bm r'} + \left(\hat{\Delta}^{0}_{\bm r, \bm r'} \right)^{\dagger}\hat{\Delta}^{0}_{\bm r, \bm r'} \right] 
    + \frac{J}{16} \sum_{a=x,y,z} \left[   \left(\hat{\rho}^{a}_{\bm r, \bm r'} \right)^{\dagger}\hat{\rho}^{a}_{\bm r, \bm r'} + \left(\hat{\Delta}^{a}_{\bm r, \bm r'} \right)^{\dagger}\hat{\Delta}^{a}_{\bm r, \bm r'} \right].
\end{equation}

% The Kitaev interaction on the $\gamma$-bond is written as
% \begin{equation}
%     K S_{i}^{\gamma}S_j^{\gamma} = - \frac{K}{2} \left[ \left(\hat{\rho}^{s}_{\bm r, \bm r'} \right)^{\dagger}\hat{\rho}^{s}_{\bm r, \bm r'}
%     + \left(\hat{\rho}^{\gamma}_{\bm r, \bm r'} \right)^{\dagger}\hat{\rho}^{\gamma}_{\bm r, \bm r'}
%     - \sum_{\gamma' \neq \gamma} \left(\hat{\rho}^{\gamma'}_{\bm r, \bm r'} \right)^{\dagger}\hat{\rho}^{\gamma'}_{\bm r, \bm r'} 
%     + \left(\hat{\Delta}^{s}_{\bm r, \bm r'} \right)^{\dagger}\hat{\Delta}^{s}_{\bm r, \bm r'}
%     + \left(\hat{\Delta}^{\gamma}_{\bm r, \bm r'} \right)^{\dagger}\hat{\Delta}^{\gamma}_{\bm r, \bm r'}
%     - \sum_{\gamma' \neq \gamma} \left(\hat{\Delta}^{\gamma'}_{\bm r, \bm r'} \right)^{\dagger}\hat{\Delta}^{\gamma'}_{\bm r, \bm r'} 
%     \right].
% \end{equation}

Lastly, we write the symmetric off-diagonal exchange interaction in terms of these fermion bilinears. For the $z$-bond, the interaction is given by:
\begin{equation}
    -\frac{\Gamma}{8} \left[ \left(\hat{\Delta}_{\bm r, \bm r'}^{x}\right)^{\dagger} \hat{\Delta}_{\bm r, \bm r'}^y  +  \left(\hat{\rho}_{\bm r, \bm r'}^{x}\right)^{\dagger} \hat{\rho}_{\bm r, \bm r'}^y + h.c. \right].
\end{equation}
{As before, the $x$ and $y$-bond terms can be obtained by the permutation  $x,y,z \rightarrow y,z,x$.}
% For the $x$-bond, it takes the form:
% \begin{equation}
%     \frac{\Gamma}{8} \left[ \left(\hat{\Delta}_{\bm r, \bm r'}^{y}\right)^{\dagger} \hat{\Delta}_{\bm r, \bm r'}^z  -  \left(\hat{\rho}_{\bm r, \bm r'}^{y}\right)^{\dagger} \hat{\rho}_{\bm r, \bm r'}^z + h.c. \right].
% \end{equation}
% For the $y$-bond, the interaction is:
% \begin{equation}
%     \frac{\Gamma}{8} \left[ -\left(\hat{\Delta}_{\bm r, \bm r'}^{z}\right)^{\dagger} \hat{\Delta}_{\bm r, \bm r'}^x  -  \left(\hat{\rho}_{\bm r, \bm r'}^{z}\right)^{\dagger} \hat{\rho}_{\bm r, \bm r'}^x + h.c. \right].
% \end{equation}

For any pair of the above fermion bilinears $\hat{A}\hat{B}$, the mean-field decoupling
\begin{equation}
    \hat{A}\hat{B} \sim \hat{A}B+ A\hat{B} - AB,
\end{equation}
where 
\begin{align}
    A = \langle \hat A \rangle
\end{align}
is to be determined self-consistently.

In the main text, we analyzed the mean-field solution for the case with $J = \Gamma =B= 0$. 
Including these terms will renormalize the parton band structure. This normalization will modify the quasiparticle interference signal, analogous to the behavior in metals.

\section{Evaluation of the LDOS}
\subsection{Chargon LDOS}

In the main text, we arrive at the following expression for chargon LDOS, 
\begin{align}
    \rho_C(\bm r, E) = \sum_m \sum_\mu |\langle \Psi^C_m| a^\dagger_{\bm r, \mu}| \Psi^C_{0} \rangle|^2 \delta(E_m^C - E).
\end{align}
The excited states $|\Psi^C_m \rangle$ are those of $H_t$, and are written as, 
\begin{align}
    |\Psi^C_m \rangle = a^\dagger_m |\Psi^C_0 \rangle,
\end{align}
where 
\begin{align}
    a^\dagger_m = U_{\bm r,m} a^\dagger_{\bm r}.
\end{align}
We drop the $\mu$ index for now, as different species of chargons do not mix. 
With the matrix elements $W_{\bm r, \bm r'}$ evaluated, we can numerically calculate $U_{\bm r, m}$ by exactly diagonalizing the Hamiltonian. 
After obtaining $U_{\bm r, m}$ the chargon LDOS takes the following expression, 
\begin{align}
     \rho_C(\bm r, E) = 2 \sum_m |U_{\bm r, m}|^2 \delta(E_m^C - E),
\end{align}
with the factor of $2$ from the sum over $\mu$. 
In numerically evaluating this formula on a finite system size of linear size $L$ the energy level spacing goes to zero as $1/L$. 

In all of our numerical computations, we use the Gaussian broadening
\begin{align}
    \delta(E_m^C - E) \approx \frac{1}{\sqrt{2\pi \sigma^2}} \exp\left[-\frac{(E_m^C - E)^2}{2\sigma^2}\right]. 
\end{align}
This is similar to adding a non-zero inverse lifetime to the quasiparticles. 
The standard deviation $\sigma$ is chosen to be of the order of the level spacing and such that the DOS is smooth. 

\subsection{Spinon LDOS}
In the main text, we arrive at the following expression for spinon LDOS, 
\begin{align}
    \rho_S(\bm r, E) =\sum_n \bigg[|\langle \Psi_n^S |c_{\bm r} | \Psi_0^S \rangle |^2   + \sum_\alpha |\langle \Psi_n^S |b^\alpha_{\bm r} | \Psi_0^S \rangle |^2 \bigg] \delta(E_m^S - E). 
\end{align}
The operator $c_{\bm r}$ does not change the vison configuration of $|\Psi_0^S \rangle$, and the evaluation of these terms requires only knowing the matrix $Q$ in Eq.~(\ref{eq:off_diag_decomp}). 
From Eq.~(\ref{eq:diag_kitaev_ham})  we see that the excited states contributing to a nonzero overlap of $\langle \Psi_n^S |c_{\bm r} | \Psi_0^S \rangle$ take the form 
\begin{align}
    |\Psi^S_n\rangle =  \psi_n^\dagger |\Psi^S_0\rangle, 
\end{align}
where $\Psi_0^S$ is defined as the state belonging to the vison-free subspace and the vacuum of $\psi_m$. 
We write $c_{\bm r} = Q_{\bm r, i} b_i = Q_{\bm r, 2n} b_{2n} + Q_{\bm r, 2n+1} b_{2n + 1}$, and thus, 
\begin{align}
    \langle \Psi_n^S |c_{\bm r} | \Psi_0^S \rangle = Q_{\bm r, 2n} - \frac{1}{i} Q_{\bm r, 2n + 1} \rightarrow  |\langle \Psi_n^S |c_{\bm r} | \Psi_0^S \rangle|^2 = Q_{\bm r, 2n}^2 + Q_{\bm r, 2n + 1}^2. 
\end{align}

The evaluation of the overlap $\langle \Psi_n^S |b^\alpha_{\bm r} | \Psi_0^S \rangle $ is more involved since the operator $b^\alpha_{\bm r}$ excites two neighboring visons at the plaquettes sharing the $\alpha$-edge connected to the site $\bm r$. 
The full ground state of the Kitaev model takes a tensor product form, 
\begin{align}
    | \Psi^S_0 \rangle = | W_0 \rangle \otimes | \Phi_0 \rangle 
\end{align}
where $| W_0 \rangle $ describes the vison configuration, and $| \Phi_0 \rangle $ describes the matter fermion $c_{\bm r}$ as discussed above. 
The $b^\alpha_{\bm r}$ operator acts on the vison subspace, 
\begin{align}
    b^\alpha_{\bm r} | \Psi^S_0 \rangle = | W' \rangle \otimes | \Phi_0 \rangle 
\end{align}
with $| W'\rangle $ describing the state with two visons excited. 
To obtain a nonzero overlap $\langle \Psi_n^S |b^\alpha_{\bm r} | \Psi_0^S \rangle$, the state $| \Psi_n^S  \rangle $ must have the same vison configuration as $|b^\alpha_{\bm r} | \Psi_0^S \rangle$. We can always choose an excited state with a matching vison configuration, so we write
\begin{align}
    | \Psi^S_n \rangle = | W' \rangle \otimes | \Phi'_n \rangle. 
\end{align}
and obtain 
\begin{align}
    \langle \Psi_n^S |b^\alpha_{\bm r} | \Psi_0^S \rangle = \langle \Phi'_n | \Phi_0 \rangle. 
\end{align}

There are multiple possibilities for the state $|\Phi'_n \rangle$, the one with the lowest energy being the ground state in the subspace of the excited visons. 
We label this state as $|\Phi'_0 \rangle$ and focus now on the evaluation of $\langle \Phi'_0 | \Phi_0 \rangle$. 
Similar to how the state $| \Phi_0 \rangle $ is defined as the state annihilated by $\psi_m$, the state  $| \Phi'_0 \rangle$ is defined as the state annihilated by a different set of operators $\psi'_m$. 
Thus, we are asking about the overlap between two different Bogoliubov vacua. 
This is the same as asking about the overlap between two superconducting states. 
This overlap is given using what is known in the nuclear physics literature as the ``Onishi" formula. 
We use a more practical formula developed in Ref.~\cite{Bertsch_2012} which gives the overlap in terms of the Pfaffian, 
\begin{align}
    \langle \Phi'_0 | \Phi_0 \rangle = \frac{e^{i\theta}}{\prod_n v'_n \prod_m v_m} \  \text{Pf} \begin{bmatrix} V^TU & V^TV'^* \\ -V'^\dagger V & U'^T V'^* \end{bmatrix} 
\end{align}
where 
\begin{align}
    \begin{bmatrix} \bm g \\ \bm g^\dagger \end{bmatrix} = \begin{bmatrix} U & V^* \\ V & U^* \end{bmatrix} \begin{bmatrix} \bm \psi \\ \bm \psi^\dagger \end{bmatrix}, \qquad g_m = \frac{1}{2} (c_{2i} + c_{2i+1})
\end{align}
and $v_m$ are the singular values of $V$. The matrices $U'$, and $V'$ are defined similarly for $\psi'_m$. 

The Pfaffian formula above is numerically challenging to compute when either $V$ or $V'$ has zero singular values. 
To circumvent this issue, we use the development in Ref.~\cite{Carlsson_2021} in which we remove the zero singular values using a Bloch-Messiah decomposition of the matrices $U$ and $V$, and $U'$ and $V'$. 
We do not spend effort extracting the exact phase $e^{i\theta}$ since for our problem we are only interested in the magnitude of the overlap. 

\begin{figure}
    \centering
    \subfloat[$t = 10^4 K$]{\includegraphics[scale=0.6,trim = 11 10 11 0, clip]{total_ldos_013k.pdf}}
    \subfloat[$t = K$]{\includegraphics[scale=0.6,trim = 11 10 11 0, clip]{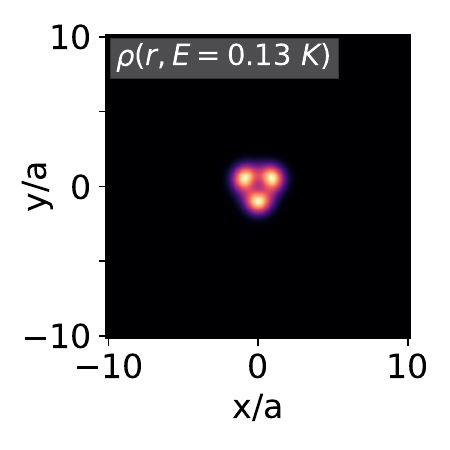}}
    \subfloat[$t = 10^4 K$]{\includegraphics[scale=0.6,trim = 11 10 11 0, clip]{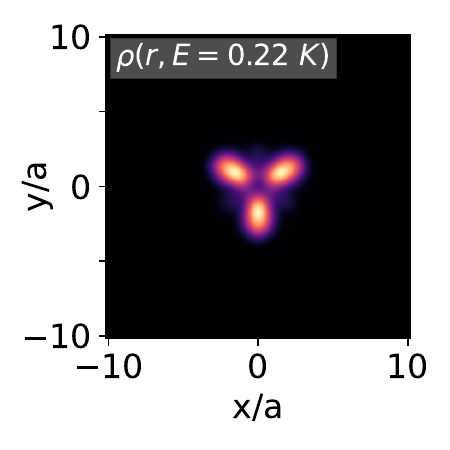}}
    \subfloat[$t = K$]{\includegraphics[scale=0.6,trim = 11 10 11 0, clip]{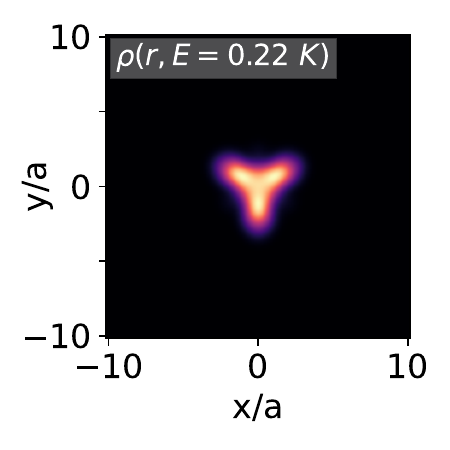}}
    
    \caption{Electron LDOS near a defect at energies below the energy for exciting a pair of visons $E_V = 0.27K$ in a clean system. We plot electron LDOS for (a) $E = 0.13 K$ and $t = 10^4 K$, (b) $E = 0.13 K$ and $t = K$, (c) $E = 0.22 K$ and $t = 10^4 K$, and (d) $E = 0.22 K$ and $t = K$ (d). All plots show that the majority of the weight is near the defect as a result of the flat vison band.}
    \label{fig:vison_effect}
\end{figure}

In principle, the computation might produce $| \Phi_0 \rangle$ and $| \Phi'_0 \rangle$ that have different fermion parity, leading to a zero overlap. 
This can always be remedied for in the case of a vacancy where the model contains dangling $b^\alpha$ Majorana operators around the vacancy. 
These can be combined into a complex fermion that is completely decoupled from the system, and we are free to choose its filling to make the parity of $| \Phi_0 \rangle$ and $| \Phi'_0 \rangle$ match.

Every evaluation of the Pfaffian formula incurs a computational cost that is $O(N^3)$ as we solve for $U'$ and $V'$. 
In our problem, we need to evaluate this Pfaffian $O(N)$ times corresponding to different locations of the vison excitations on the lattice. 
In total, the computational cost of evaluating the overlaps $\langle \Phi'_0 | \Phi_0 \rangle$ for all possible excitations is $O(N^4)$. 
Our calculations show that $|\langle \Phi'_0 | \Phi_0 \rangle|^2 \approx 0.77$ for the two neighboring visons excited by $b^\alpha_{\bm r}$.

The next task is to compute the overlap for $|\Phi'_{n} \rangle$ which represents Bogoliubov quasiparticles excited on top of the vacuum $|\Phi'_{0} \rangle$. 
These Bogoliubov excitations must come in pairs to preserve the fermion parity to have a nonzero overlap. 
In general, this can be done by writing $| \Phi'_n \rangle$ to be the vacuum of $\psi'^\dagger_0, \psi'^\dagger_1, \psi_3, \psi_4, \dots $ where $\psi'^\dagger_0, \psi'^\dagger_1$ are the two excited Bogoliubov quasiparticles. 
The computational cost of doing this evaluation is $O(N^6)$; $O(N^3)$ for evaluation of the Pfaffian, $O(N)$ places to excite the visons and, $O(N^2)$ ways to excite two Bogoliubov quasiparticles. 
{In our study, we drop these contributions, which are small since the overlap $|\langle \Phi'_0 | \Phi_0 \rangle|^2$ is large.} 

\begin{figure}
    \centering
    \subfloat[Electron LDOS]{\includegraphics[scale=0.45]{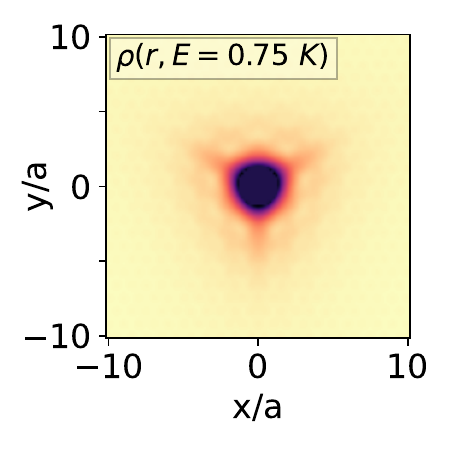}}
    \subfloat[Differentiated electron LDOS]{\includegraphics[scale=0.45]{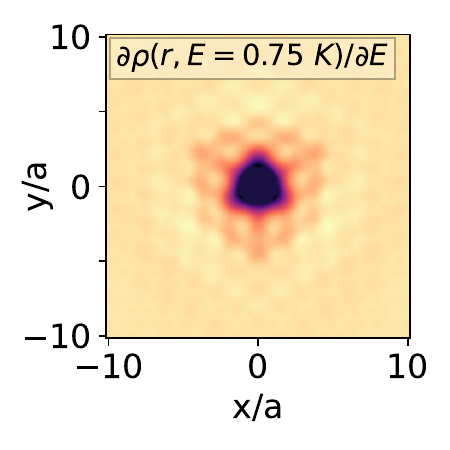}}
    \subfloat[Spinon LDOS]{\includegraphics[scale=0.45]{spinon_ldos_dirac.pdf}}

    \subfloat[]{\includegraphics[scale=0.5]{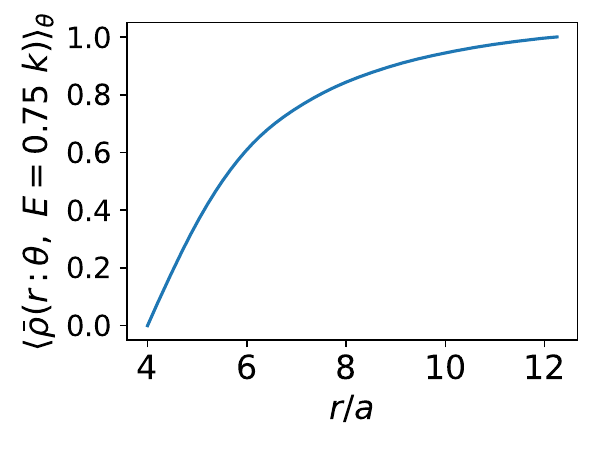}}
    \subfloat[]{\includegraphics[scale=0.5]{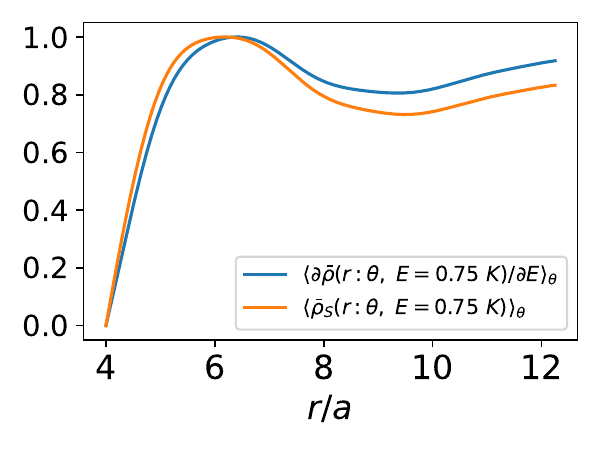}}

    \caption{Results for QPI near the spinon Dirac point at $t = 10^{4} K$. The electron LDOS looks featureless away from the defect (a), and (d). However, as shown in (b), taking a derivative with respect to energy reveals the QPI of the spinon. As comparison, the spinon LDOS obtained directly from the model [Eq. (6) in the main text] is displayed in (c). Panel (e) shows a line cut of derivative of the electron LDOS and the spinon LDOS obtained directly from the model on the same plot for comparison.}
    \label{fig:diff_spinon_dirac}
\end{figure}

\section{Numerical results for different parameters of the model}
\subsection{{QPI due to vison}}
Vison excitations near a defect cost less energy than in the bulk. 
This causes the electron LDOS to be ultra-localized near the defect for energies smaller than the bulk energy to excite a pair of visons. 
In Fig.~\ref{fig:vison_effect} we plot electron LDOS at $E = 0.13 K$ for (a) $t = 10^4 K$ and (b) $t=K$, where {QPI is mainly contributed by vison}. 
Interestingly, the QPI due to vison is insensitive to the ratio $t/K$.
In Figs.~\ref{fig:vison_effect} (c) and (d) we plot the same but for $E = 0.22K$, where again the contribution of the vison dominates. 
At this energy, the exact three-lobed shape changes slightly, but the majority of the weight is near the defect in both cases.

\subsection{\texorpdfstring{Near Dirac cone at $t=10^4 K$}{}}

In the main text, we show results for extracting the spinon dispersion near the top of the band, where the dispersion is quadratic. 
The protocol can be applied at other values of the spinon energy, particularly near the Dirac point. 
In Fig.~\ref{fig:diff_spinon_dirac} we show results of using the same protocol near the Dirac cone. 
Similar to before, we see that even though the electron LDOS looks featureless [Figs.~\ref{fig:diff_spinon_dirac} (a), and (d)], taking a derivative with respect to energy [Figs.~\ref{fig:diff_spinon_dirac} (b)], reveals the spinon LDOS [Figs.~\ref{fig:diff_spinon_dirac} (c)]. 
From Fig.~\ref{fig:diff_spinon_dirac} (e) we see that the differentiated electron LDOS and spinon LDOS agree well. 

\begin{figure}
    \centering
    \begin{minipage}{0.47\textwidth}
    \subfloat[]{\includegraphics[scale=0.41,trim = 11 10 11 0, clip]{total_ldos_5k_teqk.pdf}}
    % \subfloat[]{\includegraphics[scale=0.41,trim = 11 10 11 0, clip]{figures/diff_ldos_quadratic_teqk.pdf}}
    \subfloat[]{\includegraphics[scale=0.41,trim = 11 10 11 0, clip]{spinon_ldos_quadratic.pdf}}

    \subfloat[]{\includegraphics[scale=0.55]{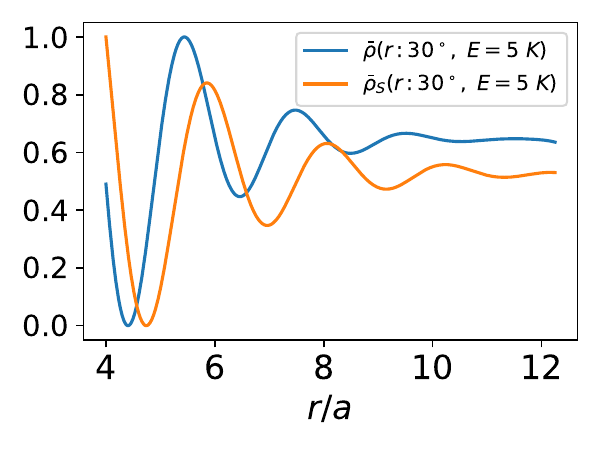}}
    \end{minipage}
    \begin{minipage}{0.47\textwidth}
    \subfloat[]{\includegraphics[scale=0.41,trim = 11 10 11 0, clip]{total_ldos_dirac_teqk.pdf}}
    % \subfloat[]{\includegraphics[scale=0.41,trim = 11 10 11 0, clip]{figures/diff_ldos_dirac_teqk.pdf}}
    \subfloat[]{\includegraphics[scale=0.41,trim = 11 10 11 0, clip]{spinon_ldos_dirac.pdf}}

    \subfloat[]{\includegraphics[scale=0.55]{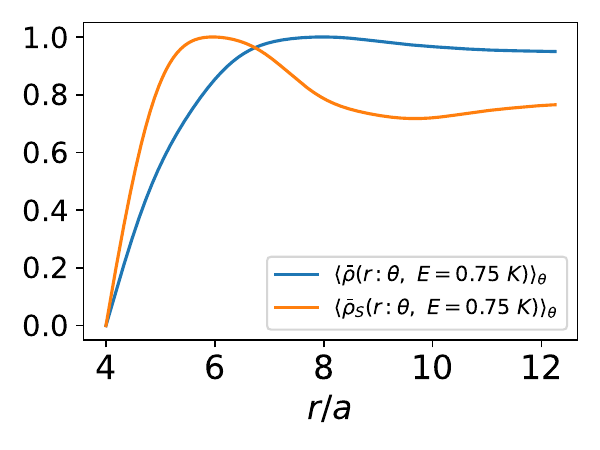}}

    \end{minipage}
    \caption{{More results with the parameters $t = K $. Panel (a) shows electron LDOS and (b) spinon LDOS calculated directly from the model [Eq. (6) in the main text] for comparison all at $E = 5.0 K$. Visually, they all look similar. Taking a line cut and plotting them on the same graph (c) we see that there is an overall agreement between the electron LDOS and the spinon LDOS. (d)--(f) are the same but for $E = 0.75K$. In this case, there is no direct correspondence between the electron LDOS to the spinon LDOS due to the lack of the separation of electron and spinon energy scale. In (c) the electron LDOS and spinon LDOS are taken along the direction of 30 degrees with respect to the $x$ axis, while in (f), the electron LDOS and spinon LDOS are averaged over all angle around the vacancy.}}
    \label{fig:more_results_teqk}
\end{figure}

\subsection{\texorpdfstring{Results for QPI at $t=K$}{}}
In Fig.~\ref{fig:more_results_teqk} we show more results for $t = K$. 
Remember that $t$ is the electron hopping strength. 
Using our mean-field treatment, we find that the chargon hopping is strongly renormalized and has a much smaller strength at $0.059t$.
This means that even when $t = K$, the spinon bandwidth is much larger than that of the chargon. 
This makes the situation similar to the case where $t \gg K$, {and we should expect electron LDOS to resemble spinon LDOS when $E\gg K$. 
This is indeed the case as shown in Figs.~\ref{fig:more_results_teqk} (a), (b), and (c) for an energy close to the top of the spinon band. Looking at Fig.~\ref{fig:more_results_teqk} (c) we see that there is reasonable agreement between the electron LDOS and spinon LDOS, except for an overall shift of the curves in space.
This agreement is expected to get better the slower the chargon is.
For $E\sim K$, as shown in Figs.~\ref{fig:more_results_teqk} (d), (e) and (f), the correspondence between the electron LDOS and spinon LDOS is less clear. This is because the bias energy is not sufficient to sample the whole spinon band, and there is no clear separation between spinon and chargon energy. The electron LDOS depends on the convolution of the spinon and chargon LDOS. Neverltess, clear spatial oscillation in electron LDOS is visible.  }

}

\end{document}